\documentstyle[12pt]{article}
\begin{document}

\title{{\Large {\bf The computation of the Conformal Killing Vectors of an $1+(n-1)$
decomposable metric}}}
\author{{Pantelis S. Apostolopoulos and Michael Tsamparlis}}
\maketitle

\section{Introduction}

A Conformal Killing Vector (CKV) $\xi ^a$ with conformal factor $\psi $ of a
general metric $g_{ab}$ is defined by the equation\footnote{%
Latin indices take the values $1,2,...,n$ ; Greek indices take the values $%
2,3,...,n$. The signature of the metric is arbitrary. Round and squared
brackets enclosing indices denote symmetric and antisymmetric part
respectively. Numbers in brackets refer to the references at the end of the
paper.}:

\begin{equation}
{\cal L}_{{\bf \xi }}g_{ab}=\xi _{a;b}+\xi _{b;a}=2\psi g_{ab}  \label{sx1}
\end{equation}
where a semicolon denotes covariant differentiation with respect to the
metric $g_{ab}$.

If $\psi =0$ the CKV reduces to a Killing Vector (KV) and if $\psi =$const.$%
\neq 0$ the CKV reduces to a Homothetic Killing Vector (HKV). The
determination of the CKVs of a general metric is important because, among
others, it allows the simplification of the metric and consequently the
easier study of the geometry of the space. For example if $\xi ^a$ is a non
null gradient CKV i.e $\xi ^a=\phi ^{,a}$ then \cite{Petrov} there exists a
coordinate system $\{{x^1,x^\alpha \}}$ such that $\phi ^{,a}=\delta _1^a,$ $%
\psi =\psi (x^1)$ and :

\begin{equation}
g_{ab}=g_{11}(dx^1)^2+\frac 1{g_{11}}\Gamma _{\alpha \beta
}(x^2,x^3,..,x^n)dx^\alpha dx^\beta  \label{sx1a}
\end{equation}
where:

\begin{equation}
g_{11}=\frac 1{2\int \psi (x^1)dx^1+C}  \label{sx1b}
\end{equation}
and $\Gamma _{\alpha \beta }(x^2,x^3,..,x^n)$ are smooth functions on their
arguments. A recent generalization of this result to non gradient non-null
CKVs has appeared in \cite{Tsamp1}.

As a roule, the determination of the CKVs of a metric is via the solution of
the differential equations (\ref{sx1}) which is not always an easy or
possible task. Thus it is important to developed general theorems which for
general classes of metrics will allow the computation of the CKVs without
the explicit solution of the differential equations. In the present paper
one such method is developed for the calculation of the CKVs of a general $%
1+(n-1)$ decomposable metric in terms of the Killing vectors and the
gradient CKVs of the associated $(n-1)$ metric.

\section{The CKVs of a $1+(n-1)$ metric}

A metric $g_{ab}$ is called $1+(n-1)$ globally decomposable if it admits a
non-null covariantly constant vector field $M^a$ say i.e. $M_{a;b}=0$. It
follows that $M^a$ is a gradient KV hence Petrov's statement applies which
means that there exist coordinates $\{x^1,x^\alpha \}$ in which ${\bf M}%
=\partial x^1$ and the metric is written:

\begin{equation}
ds^2=\epsilon ({\bf M})(dx^1)^2+g_{\alpha \beta }(x^\gamma )dx^\alpha
dx^\beta  \label{sx3}
\end{equation}
$\epsilon ({\bf M})=M^2$ is the sign of $M^a$ and the quantities $g_{\alpha
\beta }(x^\gamma )$ are the components of the metric of the $(n-1)$ space $%
x^1=$const.

Many well known and important metrics, especially in General Relativity, are
globally decomposable or conformally related to a globally decomposable
spacetime. For example such classes of metrics are the
Friedmann-Robertson-Walker metric $ds^2=-dt^2+R^2(t)\left[ \left( \frac
1{1+\frac \varepsilon 4{\bf x}^2}\right) ^2(dx^2+dy^2+dz^2)\right] $, the
G\"{o}del-type metrics \cite{RT,ART} $ds^2=-[dt+H(r)d\Phi
]^2+dr^2+D^2(r)d\Phi ^2+dz^2$ which contain the well known G\"{o}del metric
etc. The main result of this paper is the following Theorem:

\noindent {\bf Theorem 2.1.} {\it All proper CKVs }$X^a${\it \ of an }$%
1+(n-1)${\it \ (}$n\geq 3${\it ) metric }$g_{ab}${\it \ are of the form}

\begin{equation}
X^a=f(x^a)\partial x^1+K^a  \label{sx4}
\end{equation}
{\it where\ }$f(x^a)${\it \ is a smooth function and }$K^\alpha ${\it \ are
CKVs of the }$(n-1)${\it -metric }$g_{\alpha \beta }${\it \ of the form: } 
\begin{equation}
K^\alpha =\frac 1pm(x^1)\xi ^\alpha +L^\alpha (x^\beta )  \label{sx5}
\end{equation}
{\it such that:}

{\it \noindent (a) }$\xi _\alpha =A_{,\alpha }${\it \ is a gradient CKV of
the }$(n-1)${\it -metric }$g_{\alpha \beta }${\it \ whose conformal factor }$%
\lambda ({\bf \xi })$ {\it satisfies the relation}

\begin{equation}
\lambda ({\bf \xi })_{|\alpha \beta }=p\lambda ({\bf \xi })g_{\alpha \beta }
\label{sx6}
\end{equation}
$p${\it \ being a non-vanishing constant.}

{\it \noindent (b) the function }$m(x^1)${\it \ satisfies the equation }$%
\stackrel{**}{m}+\epsilon pm=0${\it \ where ''*'' denotes differentiation
with respect to the co-ordinate }$x^1${\it \ and }$\epsilon ${\it \ is the
sign of the gradient KV which decomposes the spacetime. Furthermore the
function }$f(x^a)$ {\it is defined by the equation }$f(x^a)=-\frac \epsilon p%
\stackrel{*}{m}\lambda ({\bf \xi })+Bx^1$.

{\it \noindent (c) }$L_\alpha ${\it \ is a KV or a HKV of the }$(n-1)${\it %
-metric }$g_{\alpha \beta }${\it \ which is not a gradient vector field and
its bivector equals }$F_{ab}({\bf K})${\it . Then the non-gradient KVs of
the }$(n-1)${\it -metric are identical with those of the }$n${\it -metric
and the HKVs of the }$1+(n-1)${\it \ metric are of the form }$Bx^1\partial
_1+L^\alpha \partial _\alpha ${\it \ where }$L^\alpha ${\it \ is HKV of the }%
$(n-1)${\it -metric with conformal factor }$B${\it . }$\Box $

\noindent {\bf Proof}

\noindent Suppose that $X^a$ is a general (smooth) vector field of the $%
1+(n-1)$-metric $g_{ab}$. We decompose $X^a$ along and normally to $M^a$:

\begin{equation}
X_a=f(x^b)M_a+X_a^{\prime }  \label{sx7}
\end{equation}
where $X_a^{\prime }=h_a^bX_b$, $h_{ab}=g_{ab}-\epsilon ({\bf M})M_aM_b$ is
the projection tensor and $f(x^a)$ is a smooth function. We define the
vector $K_\alpha $ in the $(n-1)$-space $x^1=$const. by the requirement $%
X_a^{\prime }=K_\alpha \delta _a^\alpha $ so that:

\begin{equation}
X_a=f(x^b)\zeta _a+K_\alpha \delta _a^\alpha .  \label{sx8}
\end{equation}
Taking the covariant derivative of the smooth vector field $X^a$ it is
always possible to write:

\begin{equation}
X_{a;b}=\psi ({\bf X})g_{ab}+H_{ab}({\bf X})+F_{ab}({\bf X})  \label{sx9}
\end{equation}
where the quantities $\psi ({\bf X}),H_{ab}({\bf X})$ and $F_{ab}({\bf X})$
are defined as follows:

\begin{eqnarray}
\psi ({\bf X}) &=&\frac 1nX_{;a}^a  \nonumber \\
&&  \nonumber \\
H_{ab}({\bf X}) &=&X_{(a;b)}-\psi ({\bf X})g_{ab}  \label{sx10} \\
&&  \nonumber \\
F_{ab}({\bf X}) &=&X_{[a;b]}  \nonumber
\end{eqnarray}
Combining eqs (\ref{sx8}) and (\ref{sx9}) we obtain the following relations:

\begin{equation}
4\psi ({\bf X})=\stackrel{*}{f}+3\lambda ({\bf K})  \label{sx11}
\end{equation}

\begin{equation}
H_{ab}({\bf X})=\left( 
\begin{array}{cc}
\epsilon (\stackrel{*}{f}-\psi ({\bf X})) & \frac \epsilon 2(f_{,\alpha
}+\epsilon \stackrel{*}{K_\alpha }) \\ 
\frac \epsilon 2(f_{,\alpha }+\epsilon \stackrel{*}{K_\alpha }) & {\cal H}%
_{\alpha \beta }({\bf K})+\frac 14(\lambda -\stackrel{*}{f})g_{\alpha \beta }
\end{array}
\right)  \label{sx12}
\end{equation}

\begin{equation}
F_{ab}({\bf X})=\left( 
\begin{array}{cc}
{\bf \bigcirc } & \frac \epsilon 2(f_{,\alpha }-\stackrel{*}{\epsilon
K_\alpha }) \\ 
-\frac \epsilon 2(f_{,\alpha }-\epsilon K_\alpha ) & {\cal F}_{\alpha \beta
}({\bf K})
\end{array}
\right)  \label{sx13}
\end{equation}
where the vector field $K_\alpha $ has been decomposed in a similar way:

\begin{equation}
K_{\alpha |\beta }=\lambda ({\bf K})g_{\alpha \beta }+{\cal H}_{\alpha \beta
}({\bf K})+{\cal F}_{\alpha \beta }({\bf K})  \label{sx14}
\end{equation}
and ''$|$'' denotes covariant differentiation w.r.t. $(n-1)$-metric $%
g_{\alpha \beta }$.

The requirement that $X^a$ is a CKV $H_{ab}({\bf X})=0$ implies:

\begin{equation}
\stackrel{\ast }{f}=\psi ({\bf X})  \label{sx15}
\end{equation}

\begin{equation}
f_{,\alpha }=-\epsilon \stackrel{*}{K_\alpha }  \label{sx16}
\end{equation}

\begin{equation}
{\cal H}_{\alpha \beta }({\bf K})=0\qquad \mbox{and}\qquad \psi ({\bf X}%
)=\lambda ({\bf K}).  \label{sx17}
\end{equation}
Taking into account the integrability conditions of $f$ i.e. $\stackrel{*}{f}%
_{,\alpha }=(f_{,\alpha })^{*}$ and $f_{,\alpha \beta }=f_{,\beta \alpha }$
we obtain the additional equations:

\begin{equation}
\lambda ({\bf K})_{,\alpha }=-\epsilon \stackrel{**}{K}_\alpha  \label{sx18}
\end{equation}
\begin{equation}
\stackrel{\ast }{\cal F}_{\alpha \beta }({\bf K)}=0.  \label{sx19}
\end{equation}
Consider first the KVs of the $n$-metric $g_{\alpha \beta }$. These are
defined by the condition $\psi ({\bf X})=\lambda ({\bf K})=0$ which by (\ref
{sx18}) and (\ref{sx19}) implies that $K_\alpha (x^i)=K_\alpha (x^\rho )$
otherwise the $n$-metric admits further a covariantly constant vector field
which we do not assume to be the case. Then (\ref{sx15}) and (\ref{sx16})
imply that $f=$const. and without loss of generality we may take $f=0$. We
conclude that the KVs of the $(n-1)$-metric $g_{\alpha \beta }$ are KVs of
the full metric $g_{ab}$. Furthermore $F_{ab}({\bf X})\equiv {\cal F}%
_{\alpha \beta }\delta _a^\alpha \delta _b^\beta $.

Using the same arguments we prove that the HKVs of the $n$-metric $g_{ab}$
are of the form:

\begin{equation}
Bx^1\partial _1+L^\alpha \partial _\alpha  \label{sx20}
\end{equation}
where $L^\alpha $ is HKV of the $(n-1)$-metric $g_{\alpha \beta }$ with
conformal factor $B$. It is woth noticing that the full metric $g_{ab}$
admits HKVs if and only if the $(n-1)$-metric $g_{\alpha \beta }$ admits
HKVs.

We consider next the proper CKVs of the $n$-metric. Differentiating equation
(\ref{sx18}) we get:

\begin{equation}
\lambda ({\bf K})_{,\alpha }\mid _\beta =-\epsilon \stackrel{**}{\lambda }(%
{\bf K})g_{\alpha \beta }.  \label{sx21}
\end{equation}
The form of equation (\ref{sx21}) and the decomposability of the $n$-metric
implies that we must look for solutions of the form:

\begin{equation}
\lambda ({\bf K})=m(x^1)A(x^\rho )+B(x^\rho ).  \label{sx22}
\end{equation}
Differentiating and using equation (\ref{sx18}) we find:

\begin{equation}
m(x^1)A_{,\alpha }+B_{,\alpha }=-\epsilon \stackrel{**}{K}_\alpha .
\label{sx23}
\end{equation}
Differentiating again and using equation (\ref{sx21}) we get:

\begin{equation}
m(x^1)A_{|\alpha \beta }+B_{|\alpha \beta }=-\epsilon \stackrel{**}{m}%
(x^1)A(x^\rho )g_{\alpha \beta }.  \label{sx24}
\end{equation}
Because $A(x^\rho ),B(x^\rho )$ are functions of $x^\rho $ only, equation (%
\ref{sx24}) implies that:

\begin{equation}
m(x^1)A_{|\alpha \beta }+\epsilon \stackrel{**}{m}(x^1)A(x^\rho )g_{\alpha
\beta }=C_1  \label{sx25}
\end{equation}

\begin{equation}
B_{|\alpha \beta }=-C_1  \label{sx26}
\end{equation}
where $C_1$ is a constant.

Differentiating (\ref{sx25}) w.r.t. $x^1$ we find:

\begin{equation}
A_{|\alpha \beta }=pA(x^\rho )g_{\alpha \beta }  \label{sx27}
\end{equation}

\begin{equation}
\stackrel{\ast *}{m}+\epsilon pm=C_2  \label{sx28}
\end{equation}
where $p,C_2$ are constants. Equation (\ref{sx27}) says that if $p\neq 0$
then $A_{,\alpha }$ is a gradient CKV of $g_{\alpha \beta }$ with conformal
factor $pA(x^\rho )$ and if $p=0$ then $A_{,\alpha }$ is a gradient KV.
Consequently we require $p\neq 0$. We compute easily $(A_{,\alpha
}A^{,\alpha })_{|\beta }=2pA_{,\beta }$ thus $A_{,\alpha }$ is not a null
vector field. Combining (\ref{sx27}) and (\ref{sx28}) with (\ref{sx25}) we
find $C_1=C_2=0$. Thus $B_{,\alpha }=0$ and the function $m(x^1)$ satisfies
the equation:

\begin{equation}
\stackrel{\ast *}{m}+\epsilon pm=0  \label{sx29}
\end{equation}
which means that ( $p\neq 0$) $m(x^1)=\sin (\surd px^1),\cos (\surd px^1)$
for $\epsilon p=1$ and $m(x^1)=\sinh (\surd px^1),\cosh (\surd px^1)$ for $%
\epsilon p=-1$.

Integrating (\ref{sx18}) it follows:

\begin{equation}
\stackrel{\ast }{K}_\alpha =-\epsilon \int m(x^1)dx^1A_{,\alpha }+D_\alpha
(x^\beta ).  \label{sx31}
\end{equation}
Differentiating and taking the antisymmetric part we find that $D_\alpha
(x^\beta )$ is a gradient vector field which we denote by $E_{,\alpha
}(x^\beta )$.

Replacing (\ref{sx29}) in (\ref{sx31}) and integrating we find:

\begin{equation}
K_\alpha =\frac 1pm(x^1)A_{,\alpha }+E_{,\alpha }x^1+L_\alpha (x^\beta ).
\label{sx32}
\end{equation}
Differentiating (\ref{sx32}) and using the fact that $K_\alpha $ is a CKV of 
$g_{\alpha \beta }$ with conformal factor $\lambda ({\bf K})$ we find:

\begin{equation}
E_{|\alpha \beta }=0  \label{sx33}
\end{equation}
\begin{equation}
L_{\alpha |\beta }=Bg_{\alpha \beta }+F_{\alpha \beta }({\bf K}).
\label{sx34}
\end{equation}
From (\ref{sx33}) we obtain $E_{,\alpha }=0$. Thus the proper CKV $K_\alpha $
takes the form:

\begin{equation}
K_\alpha =\frac 1pm(x^1)A_{,\alpha }+L_\alpha (x^\beta )  \label{sx35}
\end{equation}
where the vector field $L_\alpha $ is a KV or a HKV of the $(n-1)$-metric
whose bivector is equal to the bivector of $K_\alpha $. Finally integrating
eqn (\ref{sx15}) we obtain:

\begin{equation}
f(x^a)=-\frac \epsilon p\stackrel{*}{m}\lambda ({\bf \xi })+Bx^1.\mbox{ }%
\label{sx36}
\end{equation}
From Theorem 2.1 we conclude that for the computation of the proper CKVs of
a $1+(n-1)$ decomposable space one has to know only the KVs, the HKV and the
gradient CKVs of the $(n-1)$ space $x^1=$const.. This reduction in the
computation of the CKVs is essential and as we will show in the next section
in many cases allows us to compute the CKVs without solving the partial
differential equations (\ref{sx1}).

One sattle point is the case $n=3$ because then the $(n-1)$ space is two
dimentional and has an infinite dimentional conformal algebra. In this case
we use only the gradient CKVs (\ref{sx27}) of the two dimentional space
which form a closed subalgebra.

Using Theorem 2.1 we can prove that a given metric $g_{ab}$ does not admit 
{\it proper} CKVs. Indeed from Theorem 2.1 we infer that the metric $g_{ab}$
does not admit proper CKVs if the $(n-1)$ space metric $g_{\alpha \beta }$
does not admit \underline{gradient} CKVs. To find a necessary condition for
this to be the case we apply Ricci's identity to the vector defined in (\ref
{sx35}) and using eqs (\ref{sx22}),(\ref{sx27}) we obtain:

\begin{equation}
m(x^1)\left[ \lambda ({\bf \xi })_{,\gamma }g_{\alpha \beta }-\lambda ({\bf %
\xi })_{,\beta }g_{\alpha \gamma }-\frac 1pR_{\sigma \alpha \beta \gamma
}\lambda ({\bf \xi })^{,\sigma }\right] =R_{\sigma \alpha \beta \gamma
}L^\sigma -L_{\alpha |\beta \gamma }+L_{\alpha |\gamma \beta }.  \label{sx37}
\end{equation}
The rhs vanishes identically due to Ricci's identity applied to the vector $%
L^\alpha .$ Thus this equation gives the condition:

\begin{equation}
\lambda ({\bf \xi })_{,\gamma }g_{\alpha \beta }-\lambda ({\bf \xi }%
)_{,\beta }g_{\alpha \gamma }=\frac 1pR_{\sigma \alpha \beta \gamma }\lambda
({\bf \xi })^{,\sigma }.  \label{sx38}
\end{equation}
Contracting the indices $\alpha ,\gamma $ we get:

\begin{equation}
\lambda ({\bf \xi })_{,\alpha }{\bf =}-\frac 1{2p}R_{\alpha \beta }\lambda (%
{\bf \xi })^{,\beta }.  \label{sx39}
\end{equation}
This equation shows that $\lambda ({\bf \xi })_{,\alpha }$ is an eigenvector
of the Ricci tensor associated with the $(n-1)$-metric $g_{\alpha \beta }$
with non-zero eigenvalue $-\frac 1{2p}$. To express this statement in terms
of the metric only we rewrite (\ref{sx39}) as follows:

\begin{equation}
(R_{\alpha \beta }+2pg_{\alpha \beta })\lambda ({\bf \xi })^{,\beta }=0.
\label{sx40}
\end{equation}
Thus in order that the $(n-1)$-metrtic $g_{\alpha \beta }$ does not admit
proper gradient CKVs the following necessary (not sufficient) condition must
be satisfied:

\begin{equation}
\det (R_{\alpha \beta }+2pg_{\alpha \beta })=0.  \label{sx41}
\end{equation}
This condition can be applied directly to any given $1+(n-1)$-metric and
together with (\ref{sx6}) and (\ref{sx39}) provides a simple criterion on
the existence of gradient CKVs by the $(n-1)$-metric $g_{\alpha \beta }$
and, by Theorem 2.1., for the existense of proper CKVs for the $1+(n-1)$
metric $g_{ab}$.

\section{Applications}

One important class of applications of Theorem 2.1 exists when the $(n-1)$
space is a space of constant curvature. Indeed it is well known that the
metric $g_{ab}$ of a space of constant curvature is conformally related to
the flat metric $\bar{g}_{ab}$:

\begin{equation}
g_{ab}=N^2(x^c)\bar{g}_{ab}  \label{sx42}
\end{equation}
where $N(x^c)=\frac 1{1+\frac K4(x^ax_a)}$ and $K=\frac R{12},$ $R$ being
the curvature scalar of the space.

This means that these metrics have the same CKVs but with different
conformal factors and bivectors. If we denote by $\phi ({\bf X),}$ $\bar{F}%
_{ab}({\bf X})$ the conformal factor and the bivector of the generic CKV $%
X^a $ for the flat metric and $\psi ({\bf X}),$ $F_{ab}({\bf X})$ the
corresponding quantities for the metric $g_{ab}$ then the following
relations are easy to establish:

\begin{equation}
\psi ({\bf X})={\bf X}(\ln N)+\phi ({\bf X)}  \label{sx43}
\end{equation}

\begin{equation}
F_{ij}({\bf X})=N^2\bar{F}_{ij}({\bf X})-2NN_{[,i}X_{j]}.  \label{sx44}
\end{equation}
The generic CKV of the flat metric $\bar{g}_{ij}$ is:

\begin{equation}
X^i=a^i+a_{..j}^ix^j+bx^i+2({\bf b\cdot x})x^i-b^i({\bf x\cdot x})
\label{sx45}
\end{equation}
where $a^i,$ $a_{..j}^i,$ $b,$ $b^i$ are constants and $({\bf x\cdot x})=%
\overline{g}_{ij}x^ix^j$. KVs are defined by the constants $a^i,$ $a_{..j}^i$%
, there exists only one HKV defined by the constant $b$ and the constants $%
b^i$ define the remaining $n$ Special CKVs\footnote{%
A CKV is called Special Conformal Killing Vector iff the conformal factor $%
\psi $ satisfies the condition $\psi _{;ab}=0.$ The proper CKVs of flat
metrics are all Special CKVs.}.

Following standard notation \cite{Bruhat} we write for the CKVs of the
metric $g_{ab}$:

\[
{\bf P}_i=\partial _i,\qquad {\bf M}_{ij}=x_i\partial _j-x_j\partial
_i\qquad \qquad \mbox{(}n\mbox{ KVs)} 
\]
\[
{\bf H}=x^i\partial _i\qquad \qquad \qquad \qquad \qquad \qquad \quad \mbox{(%
}1\mbox{ HKV)} 
\]
\[
{\bf K}_i=2x_i{\bf H}-({\bf x\cdot x}){\bf P}_i\qquad \qquad \qquad \qquad 
\mbox{(}n(n-1)/2\mbox{ SCKVs)} 
\]
with conformal factors

\begin{eqnarray}
\psi ({\bf P}_i{\bf )} &=&-\frac{KN}2x_i,\psi ({\bf M}_{ij}{\bf )}=0 
\nonumber \\
\psi ({\bf H)} &=&1-\frac{KN}2{\bf x}^2  \label{sx46} \\
\psi ({\bf K}_i) &=&2Nx_i.  \nonumber
\end{eqnarray}
and bivectors:

\begin{eqnarray}
F_{ij}({\bf P}_r) &=&KNx_{[j}\delta _{i]}^r  \nonumber \\
&&  \nonumber \\
F_{ij}({\bf M}_{rs}) &=&N^2\delta _{ij}^{rs}-KNx_{[j}a_{i]r}x^r  \nonumber \\
&&  \label{sx48} \\
F_{ij}({\bf H}) &=&0  \nonumber \\
&&  \nonumber \\
F_{ij}({\bf K}_r) &=&-4Nx_{[i}\delta _{j]}^r  \nonumber
\end{eqnarray}
Instead of these vectors we consider the following basis of CKVs of the
metric $g_{ab}$:

\begin{equation}
{\bf P}_i+\frac K4{\bf K}_i,\mbox{ }{\bf M}_{ij}  \label{sx49}
\end{equation}

\begin{equation}
{\bf H,P}_i-\frac K4{\bf K}_i  \label{sx50}
\end{equation}
From (\ref{sx46})-(\ref{sx50}) it follows that the vectors ${\bf P}_i+\frac
K4{\bf K}_i,{\bf M}_{ij}$ are non-gradient KVs of the metric $g_{ab}$ and
that the vectors ${\bf H,P}_i-\frac K4{\bf K}_i$ are gradient proper CKVs of 
$g_{ab}$ with conformal factors:

\begin{equation}
\psi ({\bf H)}=1-\frac 12KN{\bf x}^2  \label{sx51}
\end{equation}

\begin{equation}
\psi ({\bf P}_i-\frac K4{\bf K}_i)=-KNx_i.  \label{sx52}
\end{equation}
Thus we have computed all the KVs and the proper gradient CKVs of any metric
of constant curvature. Using this and Theorem 2.1 we can compute the
conformal algebra of any $1+(n-1)$ space whose $(n-1)$ space is a space of
constant curvature.

It is worth noticing that any such space is conformally flat. The proof is
simple and has as follows. We have shown that the $(n-1)$ space admits $%
(n-1)+1=n$ gradient CKVs each of which gives $2n$ CKVs (via the two
solutions $m(x^1)$) of the decomposable metric. To these vectors we add the $%
n+\frac{n(n-1)}2=\frac{n(n+1)}2$ KVs of the $(n-1)$ metric and we have in
total $\frac{(n+1)(n+2)}2$ CKVs for the $n$-metric $g_{ab}$which means that
this metric is conformally flat \cite{ Eisenhart}.

Let us show how the above apply to an important spacetime found by Reboucas
and Texeira \cite{ART} called the ART (anti Reboucas-Tiomno) metric. This
metric is defined as follows:

\begin{equation}
ds_{ART}^2=dz^2-\cos ^2(r/a)d\tau ^2+dr^2+\sin ^2(r/a)d\phi ^2.  \label{sx53}
\end{equation}
By means of the transformation:

\begin{eqnarray}
\tilde{t} &=&2\frac{\sinh (\frac \tau a)\cos \left( \frac ra\right) }{%
1-\cosh \left( \frac \tau a\right) \cos \left( \frac ra\right) }  \nonumber
\\
\tilde{x} &=&2\frac{\sin \left( \frac ra\right) \cos (\frac \phi a)}{1-\cosh
\left( \frac \tau a\right) \cos \left( \frac ra\right) }  \label{sx54} \\
\tilde{y} &=&2\frac{\sin \left( \frac ra\right) \sin \left( \frac \phi
a\right) }{1-\cosh \left( \frac \tau a\right) \cos \left( \frac ra\right) } 
\nonumber
\end{eqnarray}
it takes the standard form:

\[
ds_{ART}^2=dz^2+\frac{a^2}{\left[ 1+\frac 14(\tilde{x}^2+\tilde{y}^2-\tilde{t%
}^2\right] ^2}\left( -d\tilde{t}^2+d\tilde{x}^2+d\tilde{y}^2\right) 
\]
i.e. it is a 1+3 metric.

It is easy to show that the 3-space $z=$const. has constant positive
curvature $R=6/a^2$ hence Theorem 2.1 applies. Using the results above and
transformation (\ref{sx54}) we compute first the KVs of the 3-metric to be 
\cite{ART}:

\begin{eqnarray}
{\bf \xi }_1 &=&\tan (\frac ra)\cos (\frac \phi a)\cosh (\frac \tau
a)\partial _\tau +\cos (\frac \phi a)\sinh (\frac \tau a)\partial _r-\cot
(\frac ra)\sin (\frac \phi a)\sinh (\frac \tau a)\partial _\phi  \nonumber \\
&&  \nonumber \\
{\bf \xi }_2 &=&\tan (\frac ra)\sin (\frac \phi a)\cosh (\frac \tau
a)\partial _\tau +\sin (\frac \phi a)\sinh (\frac \tau a)\partial _r+\cot
(\frac ra)\cos (\frac \phi a)\sinh (\frac \tau a)\partial _\phi  \nonumber \\
&&  \nonumber \\
{\bf \xi }_3 &=&\tan (\frac ra)\cos (\frac \phi a)\sinh (\frac \tau
a)\partial _\tau +\cos (\frac \phi a)\cosh (\frac \tau a)\partial _r-\cot
(\frac ra)\sin (\frac \phi a)\cosh (\frac \tau a)\partial _\phi  \nonumber \\
&&  \nonumber \\
{\bf \xi }_4 &=&\tan (\frac ra)\sin (\frac \phi a)\sinh (\frac \tau
a)\partial _\tau +\sin (\frac \phi a)\cosh (\frac \tau a)\partial _r+\cot
(\frac ra)\cos (\frac \phi a)\cosh (\frac \tau a)\partial _\phi  \nonumber \\
&&  \nonumber \\
{\bf \xi }_5 &=&\partial _\tau \qquad {\bf \xi }_6=\partial _\phi \qquad 
{\bf \xi }_7=\partial _z  \label{sx55}
\end{eqnarray}
In order to determine the four gradient CKVs first we compute the gradient
CKV ${\bf H}$. Using (\ref{sx51}) and the transformation equations (\ref
{sx54}) we find:

\begin{eqnarray}
{\bf H} &=&a\left[ \frac{\sinh (\frac \tau a)}{\cos (\frac ra)}\partial
_\tau +\cosh (\frac \tau a)\sin (\frac ra)\partial _r\right]  \nonumber \\
&&  \label{sx56} \\
\psi ({\bf H)} &=&-\cos (\frac ra)\cosh (\frac \tau a)  \nonumber
\end{eqnarray}
The rest three gradient CKVs can be determined by taking the commutators of $%
{\bf H}$ with the KVs (\ref{sx55}) (we recall that the KVs of the 3-metric
are identical with those of the 4-metric).

Finally, using Theorem 2.1., we find the following eight proper CKVs for the
ART spacetime ($k=1,2,3,4):$

\begin{eqnarray}
{\bf \xi }_{(k)\alpha } &=&-a^2A_{k,\alpha }\qquad \qquad {\bf \xi }%
_{(k)3}=a^2A_{k,3}  \nonumber \\
&&  \label{sx57} \\
\psi \left[ {\bf \xi }_{(k)}\right] &=&A_k  \nonumber
\end{eqnarray}

\begin{eqnarray}
{\bf \xi }_{(k+4)\alpha } &=&-a^2B_{k,\alpha }\qquad \qquad {\bf \xi }%
_{(k+4)3}=a^2B_{k,3}  \nonumber \\
&&  \label{sx58} \\
\psi \left[ {\bf \xi }_{(k+4)}\right] &=&B_k  \nonumber
\end{eqnarray}
where:

\begin{eqnarray}
A_k &=&\cos (\frac ra)\left\{ \cosh (\frac \tau a)\sinh (\frac za),\cosh
(\frac \tau a)\cosh (\frac za),\sinh (\frac \tau a)\sinh (\frac za),\sinh
(\frac \tau a)\cosh (\frac za)\right\}  \nonumber \\
&&  \label{sx59} \\
B_k &=&\sin (\frac ra)\left\{ \cos (\frac \phi a)\sinh (\frac za),\cos
(\frac \phi a)\cosh (\frac za),\sinh (\frac \phi a)\sinh (\frac za),\sinh
(\frac \phi a)\cosh (\frac za)\right\} .  \nonumber
\end{eqnarray}
As a second application of Theorem 2.1. we prove that the G\"{o}del metric
does not admit CKVs. It is well known \cite{Nunez} that the G\"{o}del metric
admits five KVs and does not admit HKVs \cite{Hall-da Costa} but it does or
does not admits proper CKVs. The answer is that the G\"{o}del metric does
not admit proper CKVs. To prove this we use the relation (\ref{sx39}) and
the necessary condition (\ref{sx41}).

In Cartesian coordinates the G\"{o}del metric is given by the following line
element \cite{Godel}:

\begin{equation}
ds^2=-dt^2-2e^{ax}dtdy+dx^2-\frac 12e^{2ax}dy^2+dz^2\mbox{ }  \label{sx60}
\end{equation}
where $a$ is an arbitrary constant. Thus the G\"{o}del metric is a 1+3
metric along the spacelike direction $z$ and (\ref{sx41}) applies. The Ricci
tensor for the 3-metric $g_{\alpha \beta }dx^\alpha dx^\beta
=-dt^2-2e^{ax}dtdy+dx^2-\frac 12e^{2ax}dy^2$ is calculated to be:

\begin{equation}
R_{\alpha \beta }=a^2dt^2+2a^2e^{ax}dtdy+a^2e^{2ax}dy^2.  \label{sx61}
\end{equation}
hence the determinant $\det (R_{\alpha \beta }+2pg_{\alpha \beta
})=2p^2e^{2ax}(a^2-2p).$ The determinant vanishes when $p=a^2/2>0$. For this
value of $p$ the eigenvalue equation (\ref{sx39}) gives easily that $\lambda
({\bf \xi })_{,x}=\lambda ({\bf \xi })_{,y}=\lambda ({\bf \xi })_{,t}=0$,
i.e. $\lambda ({\bf \xi })=$const. which is a contradiction. Thus the
G\"{o}del space-time does not admit proper CKVs.

\bigskip

\hspace{3.5cm}Department of Physics, Section of
Astronomy-Astrophysics-Mechanics,

\hspace{3.5cm}University of Athens, Panepistemiopolis, Athens 157 83, GREECE

\end{document}